\documentclass{bmcart}

\usepackage[utf8]{inputenc} 

\usepackage{textcomp,multirow,rotating,array}
\usepackage{algorithmic}
\usepackage[ruled,vlined]{algorithm2e}
\usepackage{placeins}
\newcommand{\rot}{\rotatebox{90}}

\startlocaldefs
\endlocaldefs

\begin{document}

\begin{frontmatter}

\begin{fmbox}
\dochead{Research}


\title{Overcoming vaccine hesitancy by multiplex social network targeting: An analysis of targeting algorithms and implications}


\author[
   noteref={n1},                        
]{\inits{MF}\fnm{Marzena} \snm{F\"ugenschuh}}
\author[
   addressref={aff2},
   corref={aff2},
   noteref={n1},
   email={Feng.Fu@dartmouth.edu}
]{\inits{FF}\fnm{Feng} \snm{Fu}}


\address[id=aff1]{
  \orgname{Berliner Hochschule für Technik},
  \street{Luxemburgerstr. 10},
  \postcode{13353}
  \city{Berlin},
  \cny{Germany}                                 
}
\address[id=aff2]{%
  \orgname{Department of Mathematics},
  \street{Dartmouth College},
  \postcode{03755}
  \city{Hanover, NH},
  \cny{USA}
}


\begin{artnotes}
\note[id=n1]{Equal contributor} 
\end{artnotes}

\end{fmbox}


\begin{abstractbox}

\begin{abstract} 
Incorporating social factors into disease prevention and control efforts is an important undertaking of behavioral epidemiology. The interplay between disease transmission and human health behaviors, such as vaccine uptake, results in complex dynamics of biological and social contagions. Maximizing intervention adoptions via network-based targeting algorithms by harnessing the power of social contagion for behavior and attitude changes largely remains a challenge. Here we address this issue by considering a multiplex network setting. Individuals are situated on two layers of networks: the disease transmission network layer and the peer influence network layer. The disease spreads through direct close contacts while vaccine views and uptake behaviors spread interpersonally within a potentially virtual network. The results of our comprehensive simulations show that network-based targeting with pro-vaccine supporters as initial seeds significantly influences vaccine adoption rates and reduces the extent of an epidemic outbreak. Network targeting interventions are much more effective by selecting individuals with a central position in the opinion network as compared to those grouped in a community or connected professionally. Our findings provide insight into network-based interventions to increase vaccine confidence and demand during an ongoing epidemic. 
\end{abstract}




\end{abstractbox}
%

\end{frontmatter}



\section*{Introduction}

Mass vaccination represents a crucial strategy in preventing and mitigating the spread of infectious diseases through the establishment of herd immunity~\cite{Jacob2000}, protecting even unvaccinated individuals. Despite its demonstrated benefits, vaccine hesitancy remains a persistent issue, compounded by challenges posed by the ongoing COVID-19 pandemic~\cite{who2019, Hudson2021, CASCINI2021, Coustasse2021,de2022impact,jentsch2021prioritising}. Heightened concerns regarding vaccine safety and efficacy have exacerbated the long-standing challenge of voluntary vaccination~\cite{Larson2013,Determann2014,Kennedy2020}. 

Of particular interest, the study of how social networks influence public health behavior and vaccine choices is a significant area of research~\cite{christakis2013social}. The dynamic spread of changing opinions on vaccination and its impact on disease outbreaks through social networks constitutes a ``dueling contagion'' process~\cite{fu2017dueling}. It is imperative to recognize that social contagions can also perpetuate non-socially optimal health behaviors. For instance, the diffusion of vaccine scares among parents through social networks has led to a decline in infant vaccination rates, resulting in a surge in childhood diseases~\cite{jansen2003measles}. Both pro- and anti-vaccine opinions are not only transmitted through personal interactions~\cite{Shaham2020} but also through digital social media platforms~\cite{Salathe2011,NAYAR2019}. A comprehensive understanding of these spreading mechanisms (see, for example, a comprehensive review in Ref.~\cite{PastorSatorras2015}) is vital to leverage the positive impacts of social contagion and mitigate its adverse effects in public health efforts~\cite{campbell2013complex,salathe2008effect}.

In recent years, the role of social factors in epidemiology has attracted growing attention~\cite{bauch2013social,NAYAR2019}. Researchers have used behavior-disease interaction models to investigate the impact of various factors, such as vaccine scares or heightened disease awareness, on vaccine compliance~\cite{bauch2005imitation,Fu2010,Cardillo2013,zhang2012rational,Bhatta2019}. We refer to Refs.~\cite{Bhatta2012,wang2016statistical,Bedson2021} for systematic reviews in this regard. Going beyond interacting diffusion of the same sort~\cite{GomesDiaz-Guilera2013,Xuan2018,Chang2019}, prior studies have considered the spread of disease, health behavior, and/or information can interact in a close feedback manner, either through the same network (single-layered)~\cite{funk2009spread} or through multilayer networks~\cite{granell2013dynamical,wang2014asymmetrically,liu2016impacts,pan2019optimal,Kahana2021}. In the latter, spread of infectious disease on one layer interacts with the diffusion of health behavior on the other~\cite{Mao2012}, or coupled with a third layer of information diffusion~\cite{Mao2014}.

Motivated by the empirical work using social network targeting for improving public health interventions~\cite{kim2015social}, here we propose and compare a number of network-based targeting algorithms that aim to maximize the influence of initial seeds who are early supporters of health intervention (e.g., vaccination). This problem is also known as influence/diffusion maximization: how to select the seed users so that the total number of triggered adopters can be maximized. It is common practice to identify and target influential individuals based on various centrality measures~\cite{Masuda2009,Zao2014,Gupta2016} in both real-life networks and social media~\cite{morone2015influence}. Despite that such targeting concept is similar to identify superspreaders and targeted vaccination~\cite{PastorSatorras2002,Cohen2003-2} in epidemiology, much is unknown about diffusion maximization on multiplex networks. In this work, we fill this theoretical gap and study a variety of targeting algorithms in a multiplex setting and assess their effectiveness of mitigating an ongoing epidemic.

Specifically, we consider a two-layer multiplex, one with a social contagion network of opinion formation and the other with a spatial network of disease transmission. The novelty of our approach is twofold. First, we consider heuristic context-dependent targeting methods explicitly using multiplex networks by coupling the dynamics of social influence and behavior changes with an ongoing epidemic. Second, aside from such interplay of concurrent disease and opinion propagation, we also consider dueling contagions of opposing views on the influence network (not just the pro-vaccine opinion but also the anti-vaccine together with neutral ones) and study how they impact targeting effectiveness. In so doing, our work represents a step forward in the field of network-based interventions, by improving our understanding of how to optimize targeting interventions with potential applications across a wide range of domains.

Our present work extends beyond our previous preliminary study~\cite{fugenschuh2023overcoming} with two key contributions. First, when measuring targeting effectiveness, we not only consider the final size of the epidemic (how the disease spreading is suppressed overall), but also the peak of the epidemic (how the targeting intervention flattens the epidemic curve). Second, we  test the robustness of our targeting algorithms against variations of modeling parameters including initial conditions, network size, and density. Our results help improve our understanding of how to optimize targeting interventions by making them more practical and scalable in the real world.

\section*{Model and Methods}

\begin{figure}[h!]
\caption{\label{fig:mx}\csentence{An example of a multiplex with an opinion (top) and disease (bottom) layer, initialized on a $15\times15$\,-\,square lattice and using default values of model parameters $m, \upsilon^{-}, \upsilon^{+}, {\cal O}_s$ and $\iota$ as listed in Table~1.}}
\includegraphics[width=0.6\textwidth]{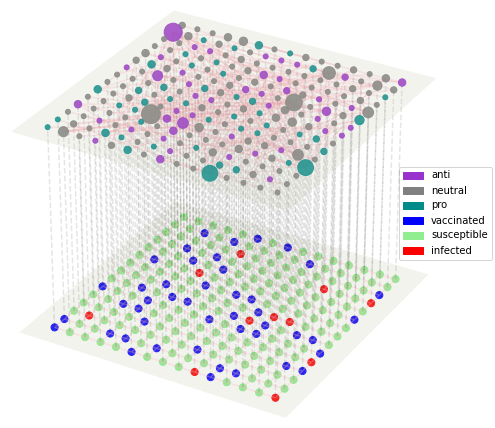}
\end{figure}

\paragraph{Model overview.}
The multiplex we study consists of two network layers, each containing the same set of individuals but with different connections between them. The first layer represents an individual's typical environment, including in-person interactions with family, coworkers, friends, and others during daily activities (including leisure time or hobbies). We assume a degree-regular network of these contacts among all individuals, and represent these relationships using a square lattice graph. On this layer, we simulate the spread of a disease. The second layer represents a virtual social network and provides a platform for opinion exchange~\cite{Agent2019}. The connections between individuals on this layer are degree-heterogeneous, generated by the Barab\'{a}si-Albert network model~\cite{barabasi1999emergence}. Such network structure allows the position of an individual to differently influence the dynamics of opinion dissemination. Past research has shown that single-layered scale-free networks promote both vaccination behavior and effective immunization~\cite{Cardillo2013}. In this work we explicitly consider a dueling contagion process~\cite{fu2017dueling} on two-layered networks as follows.

\renewcommand{\algorithmicrequire}{\textbf{Input: }}
\renewcommand{\algorithmicensure}{\textbf{Output: }}
\begin{algorithm}[htb]
\caption{\label{alg:BS} Procedure for agent-based simulations}
\begin{algorithmic}[1]
\REQUIRE All parameters listed in Table~\ref{tab:parameter}.
\STATE {Initialize the lattice graph (bottom layer) and the Barabási-Albert network (top layers).}
\STATE {Set up the opinion layer on the Barabási-Albert network}
\STATE {Set up the disease layer on the lattice graph.}
\WHILE {{\it number of infected greater than zero}}
\STATE {Select randomly an individual for update.}
\STATE {Update the states of the individual}
\IF {$rand() < \omega$} 
\STATE{Disease state updating on the disease layer.}
\ELSE 
\STATE {Opinion state and vaccine uptake on the opinion layer.}
\ENDIF
\ENDWHILE
\ENSURE The state of each individual per iteration and layer.
\end{algorithmic}
\end{algorithm}

The simulation alternates probabilistically between disease spread on one layer and opinion formation (and vaccine uptake) on the other. In each iteration, an individual is chosen at random to update their current states. The next step is to determine which layer the process should continue on. As the dynamics of the processes on the two layers are different, the probability of choosing a layer need not be equal. To this end, we introduce the parameter $\omega$ to control the relative time scales of biological and social contagions: with probability $\omega$, a state update of that individual takes place on the disease transmission layer, and otherwise with probability $1- \omega$, an opinion state update occurs on the other network layer (see Algorithm~\ref{alg:BS}). A full overview of all simulation parameters can be found in Table~\ref{tab:parameter}. Once a network layer is selected for update, the corresponding process is continued based on the individual's state in that layer (see Figure~\ref{fig:mx}), as described in further detail below.

\begin{table}[h!]
\caption{\label{tab:parameter}Default parameters used in the agent-based simulations.}
\begin{tabular}{p{0.1\textwidth}p{0.1\textwidth}p{0.1\textwidth}p{0.5\textwidth}}
\hline 
{\bf layer}  & {\bf notation} & {\bf default}  & {\bf parameter definition}  \\\hline
& $\omega$ & $0.75$  & probability of selection of the disease layer  \\
& $n$ & $50$ & size of the square lattice graph, $n^2$ is the total number of nodes in each layer\\
& $m$ & $1$ & number of edges to be attached from a new node in the Barab\'{a}si-Albert network \\
\hline
    \multirow{6}{*}{\rot{opinion}} & 
    $\upsilon^{-}$ & $10\%$ & initial percentage of anti-vaccine opinions  \\
    & $\upsilon^{+}$ & $10\%$ & initial percentage of pro-vaccine opinions  \\
    & ${\cal O}_s$ & {\it rand} & placement strategy to assign supporters  \\
    & ${\cal O}_a$ & {\it all}  & opinion adoption method   \\
    & $\rho$ & $0.25$ & probability that the contrary opinion\\
    & & & is adopted \\\hline
    \multirow{6}{*}{\rot{disease}} 
    & $\iota$ & $1\%$ & percentage of initial infection seeds \\
    & $\beta$ & $0.95$ & base probability that a susceptible gets infected\\ 
    & & & provided infections in the neighborhood \\
    & $\gamma$ & $0.25$ & probability that an infected recovers \\
    & $\eta$ & $0.02$ & probability that a susceptible individual with neutral \\
    & & & opinion gets vaccinated  \\ 
    \hline
\end{tabular}
\end{table}

\subsection*{The Opinion Contagion Layer and Network Targeting Algorithms}

Our objective is to study networks targeting methods that harness the social contagion of pro-vaccine opinions, ultimately increasing the willingness of the population to prevent an epidemic through vaccinations. Therefore, we simulate an exchange of views on the subject of vaccines. Based on the way the individuals represent their opinions, we consider a voter model with a single discrete variable with more - in our case three - states, according to the classification given in Ref.~\cite{Sirbu2017,JEDRZEJEWSKI2019}. In our setting, an individual can have one of the three discrete opinions: anti-, neutral or pro-vaccine, indicated by the values $-1,0$ and $1$ respectively. The number of supporters and opponents at the start is given by the parameters $\upsilon^{+}$ and $\upsilon^{-}$ respectively. To assess the impact of the supporters' exact positions in the network on the opinion formation process and consequentially on the epidemic, we consider the following network-based targeting methods for the initial placement of pro-vaccine opinions.

\begin{figure}[h!]
\caption{\label{fig:iniVaxx}\csentence{Network-based targeting. Shown are specific examples of initial supporter placements for each targeting method considered. Individuals are situated on the underlying spatial lattice and highlighted are only the potentially long-range connections between those supporters chosen within the opinion layer according to different targeting methods.}}
\includegraphics[width=0.9\textwidth]{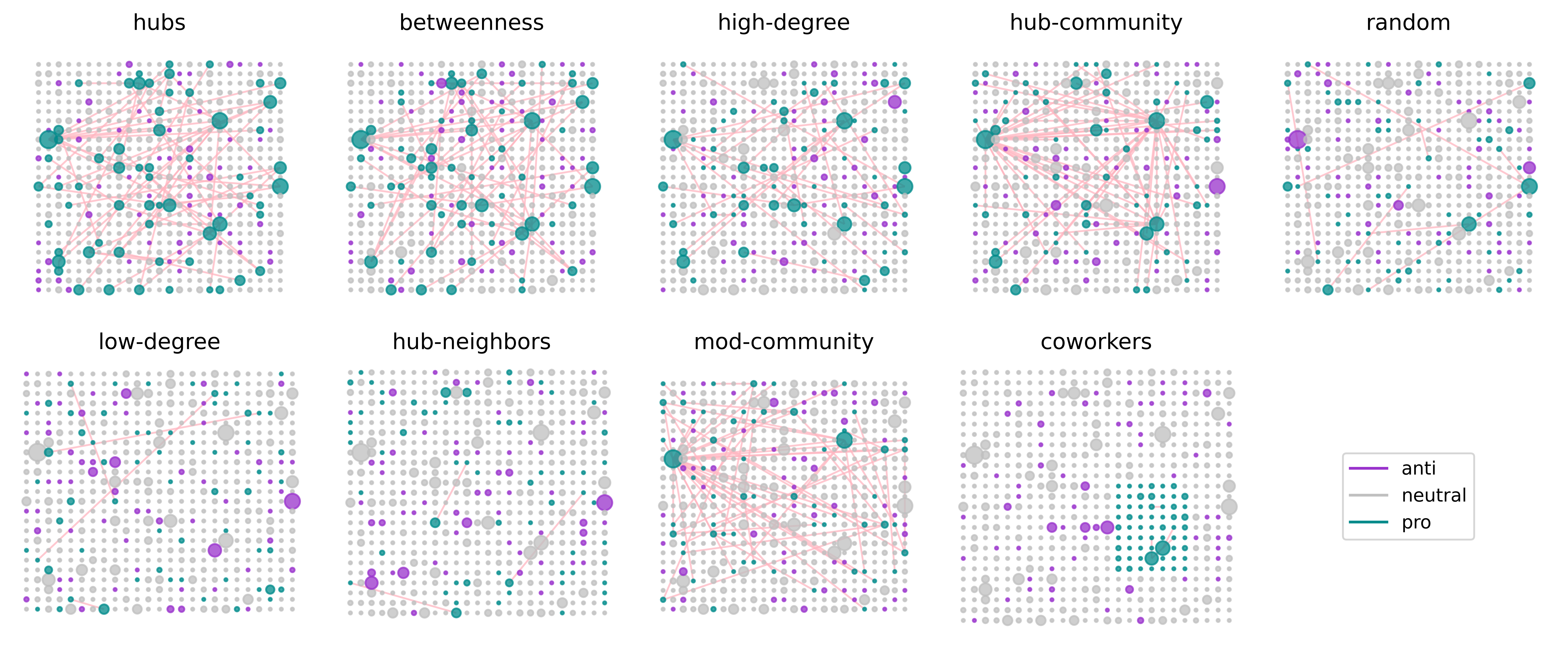}
\end{figure}

\begin{enumerate}
    \item {\it hubs}\,: The supporters are chosen from the vertices with the highest degree, referred to as hubs.
    \item {\it betweenness}\,: The probability of a vertex being chosen as a supporter is based on its betweenness centrality. The higher the betweenness value, the greater the likelihood of being selected.
    \item {\it high-degree}\,: Pro-vaccine opinions are assigned according to the degree distribution of the vertices. Again, the higher the degree, the higher the probability of being selected.
    \item {\it random}\,: Vertices are randomly selected from the entire population with a uniform probability.
    \item {\it hub-community}\,: A heuristic approach is used, starting by filling the supporter set with a vertex of the highest degree and its neighbors. Then, new vertices are repeatedly added to the supporter set by choosing neighbors of supporters with the highest degree in the opinion layer as long as they have not been chosen yet and there are places available to do so.
    \item {\it low-degree}\,: The least connected vertices have a higher probability of being chosen as vaccination supporters. Pro-vaccine opinions are assigned based on the degree distribution of the vertices, with the lower the degree, the higher the probability of being selected.
    \item {\it hub-neighbors}\,:The individuals selected are those adjacent to hubs, but not themselves hubs.
    \item {\it mod-community}\,: Supporters are members of communities determined using the Clauset-Newman-Moore greedy modularity maximization algorithm~\cite{clauset2004finding}, designed to find communities in scale-free networks, among others. Communities are selected in the order of their size until the desired number of supporters is reached.
    \item {\it coworkers}\,: Pro-vaccine opinions are assigned to vertices that form a connected rectangular-shaped sub-grid on the disease layer.
\end{enumerate}
After the initial pro-vaccine opinions are assigned, the individuals with anti-vaccine views are randomly selected from the remaining nodes. The rest of the individuals are set to neutral. We use the parameter ${\cal O}_s$ to designate the targeting method used in the simulation.

Figure~\ref{fig:iniVaxx} demonstrates how the placement of vaccine supporters varies with the different targeting methods. The network, which remains constant for all methods, is comprised of vertices arranged on a $25\times25$ lattice for the disease layer, while the opinion layer is generated using the Barab\'{a}si-Albert preferential attachment model~\cite{barabasi1999emergence} (in which each new node is connected to existing nodes with one edge, $m = 1$). Both pro- and anti-vaccine opinions are present at a proportion of $10\%$ of the total population. The sizes of vertices are scaled based on their degree and the edges connecting the vaccine supporters are displayed to give an indication of the strength of connections within the assigned pro-vaccine group. Note that all other edges have been omitted for clarity.

\paragraph{Opinion updating rule.}
After the initial views are assigned, the opinion formation process unfolds~\cite{Castellano2009}. The opinion of a selected individual is updated using one of the following five updating rules.

The focal individual adopts the opinion
\begin{enumerate}
    \item of a randomly chosen neighbour ({\it random})~\cite{Sood2005} or
    \item of the majority of the neighbors ({\it max})~\cite{Fu2008} or
    \item indicated by the sign of the sum of the opinions of the neighbors ({\it sum}) or
    \item the closest to the average of the neighboring opinions ({\it mean})~\cite{kozitsin2022formal}.    
\end{enumerate}
We also implement a hybrid method of opinion adoption that incorporates {\it all} above updating rules by randomly choosing one at each iteration step. We use the parameter ${\cal O}_a$ to specify the opinion updating rule used in the simulation.

In addition, it is often difficult for individuals to accept opposing opinions~\cite{~\cite{Deffuant2002}}, which is why we use the parameter $\rho$ to represent the probability of ultimately accepting the calculated opposing opinion. This assumption of an inertia effect when revisiting contrary opinions aligns with prior experimental observations~\cite{pitz1969inertia,traulsen2010human} and modeling studies~\cite{stark2008slower,zhang2011inertia,wang2020public}.

\subsection*{The Disease Transmission Layer}

The epidemic spreading process we consider on the disease layer is based on the Susceptible-Infected-Recovered (SIR)  model, with an extension to include the state of immunization \cite{Hens2012,Newman2002}. Individuals who opt to receive the vaccine, in particular the initial supporters in the opinion layer, once vaccinated, will remain in immunized state throughout the epidemic simulation. Infection seeds are prescribed by a predefined number of individuals in the population - given by the parameter $\iota$ - who, of course, are not vaccinated. Once the simulation is started and if it is the turn of the disease layer to update, the state of the chosen individual is updated according to one of the following rules. (At each iteration, an individual can only record at most a single state change at a given time.) 

\begin{enumerate}
    \item A susceptible pro-vaccine supporter is vaccinated without hesitancy, whereas a susceptible individual with neutral attitude will get vaccinated with a predefined very low probability $\eta$. In contrast, a susceptible anti-vaccine individual will never choose to be vaccinated. 
    \item If not vaccinated, a susceptible can get infected, provided that there are active infections in the immediate neighborhood. As a matter of fact, the more infected neighbors the higher risk of infection~\cite{PastorSatorras2015}. Thus, the parameter $\beta$ that represents the base probability of getting infected, increases by the percentage of the infections in the neighborhood: $\beta\left(1+\frac{I_u}{N_u}\right)$, where $N_u$ is the number of all neighbors of the focal susceptible $u$  and $I_u$ is the number of infectious neighbors.
    \item An infected recovers with a predefined probability $\gamma$.
    \item Once vaccinated or recovered, individuals remain in their assumed state until the end of the simulation. 
\end{enumerate}

In our simulations, we focus on exploring how the initial pro-vaccine opinion individuals should be placed on the opinion layer in order to achieve the most efficient diffusion of vaccine support views in terms of suppressing the epidemic spreading on the disease layer. Thus we mainly concentrate on varying the parameters that govern social contagion on the opinion layer, including the initial number of individuals holding pro- and anti-vaccine opinions as well as targeting methods for the pro-vaccine supporter placement and specific assumptions of individuals' view adoption. Simulation results, averaged over $100$ realizations, are shown as boxplots. Unless noted otherwise, we use default model parameter values given in Table~\ref{tab:parameter}. 

\section*{Results}

\begin{figure}[h!]
\caption{\label{fig:sim}\csentence{Impact of dueling contagions. Shown is an example of an opinion formation process (upper left) which is concurrent with a disease spread course (lower left). Such multiplex disease-behavior interactions can lead to the outgrowth of vaccination supporters and thus the suppression of the disease transmission, as  compared to scenarios having a constant rate of vaccination (upper right) or absent of any immunized individuals (lower right). All other parameters are set to be default values as given in Table~\ref{tab:parameter}.}}
\includegraphics[width=0.6\textwidth]{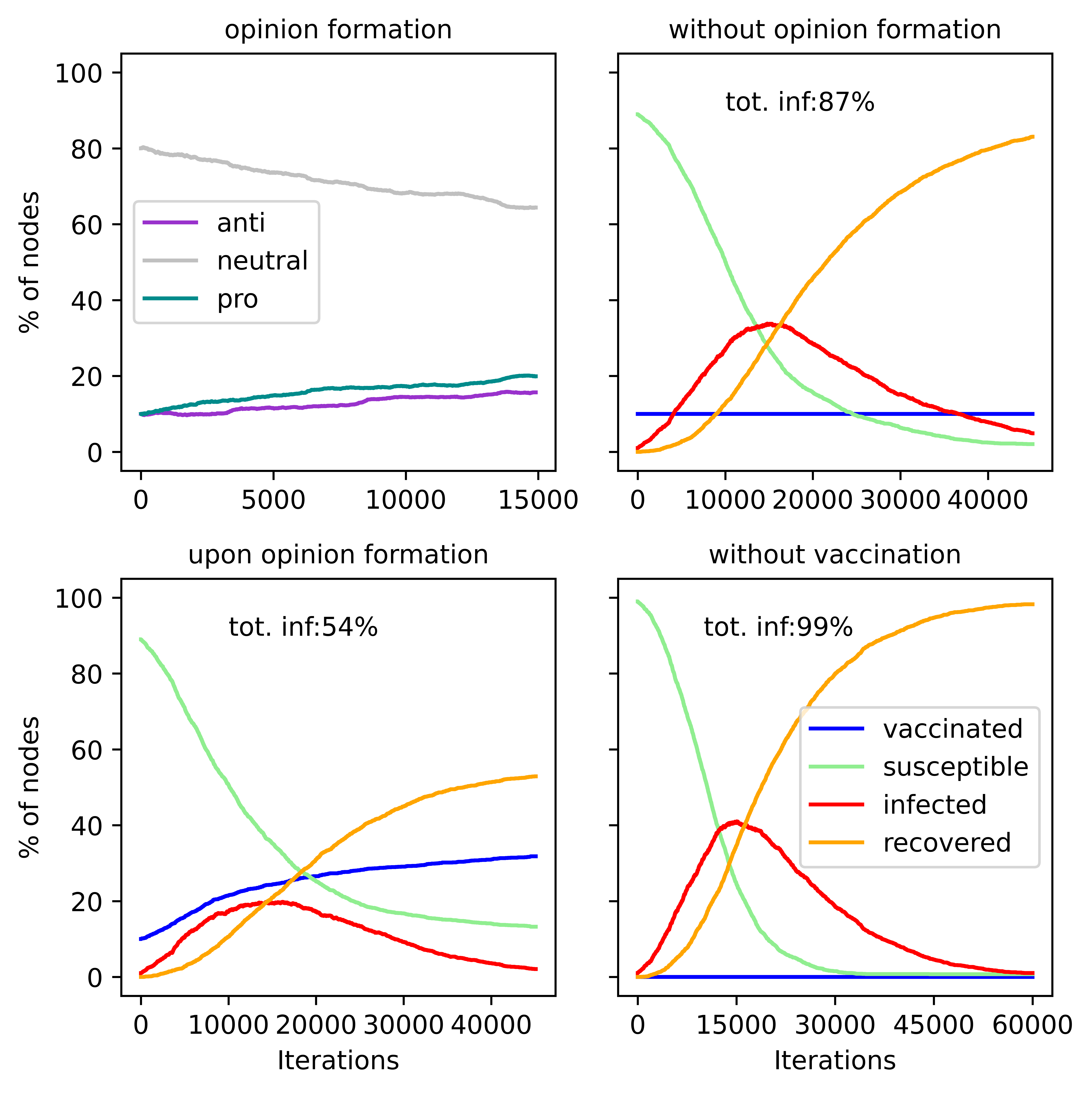}
\end{figure}

To begin with, we show the results of a typical simulation run initialized with the default values as given Table~\ref{tab:parameter}, which gives us an intuitive understanding of how the contagion process on the opinion layer can be harnessed to control disease spread (Figure~\ref{fig:sim}). On the left panels of Figure~\ref{fig:sim}, we show the simultaneous progression of both social and biological contagions -- on the opinion layer (upper left) and on the disease layer (lower left). For comparison, the panels on the right display the epidemic spread with the same initial model parameters as in the left panels but without the influence of the opinion layer: The individuals are either vaccinated only in the initial phase (upper right) or not at all (lower right). To assess the impact of the epidemic and compare various scenarios, we keep track of the number of individuals who contract the illness during each simulation run. Based on the three scenarios of disease transmission provided, we see that when the spread of the disease is suppressed by increasing vaccinations based on opinion formation, $54\%$ of the population becomes infected. On the other hand, $87\%$ of the population becomes infected when the number of immunized individuals remains constant throughout (no social contagion of vaccination), and $99\%$ if there is no vaccination at all. In the latter scenario, the epidemic lasts significantly longer, with a duration of $60,000$ iterations compared to $45,000$ iterations when vaccinations are administered.

 In evaluating the effectiveness of targeted interventions, we not only consider the final size of the epidemic~\cite{fugenschuh2023overcoming}, but also the peak as a measure of disease mitigation effort. In the following, we present a comprehensive analysis of the effectiveness of targeting algorithms, and we validate the robustness of these findings through simulations that take into account varying model parameters such as initial conditions, network size, and density. 

\subsection*{Comparison of different network targeting algorithms along with opinion updating rules}

\begin{figure}[h!]
\caption{\label{fig:iniAdopt}\csentence{Comparison of effectiveness of different targeting algorithms. Shown are the peak of infections (the maximum percentage of infections over all iterations of the epidemic spreading process) (top) and the final epidemic size (the total percentage of individuals ever infected (bottom), grouped for each targeting method of initial placement of pro-vaccine supporters (as indicated on the $x$-axis ) with respect to different opinion updating rules (corresponding to the color map in the legend). Each boxplot is based on 100 independent simulations.}}
\includegraphics[width=.9\columnwidth]{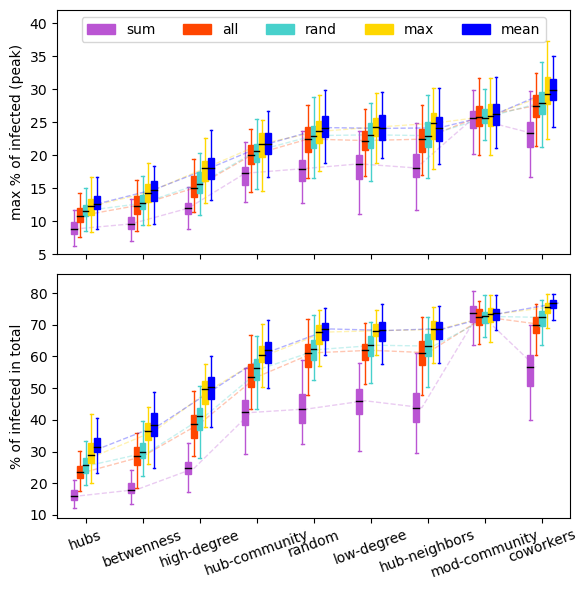}
\end{figure}

Here we investigate and compare the extent to which the spread of the disease can be impacted jointly by the initial placement of pro-vaccine supporters (${\cal O}_s$) and the specific opinion adoption method (${\cal O}_a$) applied to the opinion layer. To this end, we vary the parameter combinations governing ${\cal O}_s$ and ${\cal O}_a$ and set all others to defaults, as given in Table~\ref{tab:parameter}. 

The simulation results are presented in Figure~\ref{fig:iniAdopt}. The upper plot displays boxplots of the peak of the epidemic, indicating the maximum percentage of infected individuals during all iterations. In the lower plot, the boxplots show the final epidemic size, i.~e.~the total percentage of individuals ever infected across all iterations. Corresponding to each initial placement method of vaccine supporters (as ordered on the $x$-axis), each of the five opinion adoption methods that are grouped together is distinguished by different colors.

The evaluation of the impact of different opinion adoption methods on the spread of pro-vaccine support demonstrates that both the initial placement of vaccine supporters on the opinion layer and the chosen opinion adoption method have a crucial impact on the effectiveness of mitigating the epidemic. It is clear that the vertices with the highest network centrality values are the most influential in this social contagion process, and that the potential number of nodes that a pro-vaccine supporter can influence is a critical factor. However, when the best-connected nodes are part of a like-minded community, as seen in the {\it hub-community} and {\it mod-community} cases, their performance is similar to or even worse than the completely {\it random} approach. The density of connections between the initially placed vaccine supporters does not appear to significantly affect the spread of pro-vaccine opinions. In the case of {\it coworkers}, it is apparent that seeding a pro-vaccine opinion within a physically connected group, such as coworkers, does not result in an effective spread of pro-vaccine views. Our simulations also revealed that the scenario where low-degree vertices acting as hub neighbors could spread pro-vaccine opinions to their hub neighbors and trigger a ripple effect through the entire network is unlikely.

The assessment of the opinion adoption methods ${\cal O}_a$ reveals a consistent ordering of corresponding results (the colored boxplots in Figure~\ref{fig:iniAdopt}) per initial placement method for pro-vaccine supports, indicated by the $x$-axis. Among all five different opinion updating rules considered, the one yielding the most impact on disease mitigation appears to be taking the sign of the sum of neighboring opinions (indicated by the purple color Figure~\ref{fig:iniAdopt}). This method is followed by the {\it random} approach, which exhibits a noticeable gap in performance. The methods {\it max} and {\it mean} trail closely behind and perform almost equally with a slight disadvantage compared to the other majority approach.

The exceptional case of {\it mod-community}-based targeting highlights the limitation in exchanging or spreading opinions within a closed like-minded community, regardless of the specific opinion adoption method used. This is due to the restricted flow of information, which results in saturation when this method of targeting is applied.

\subsection*{Impact of varying initial conditions of opinion formation}

\begin{figure}[h!]
\caption{\label{fig:proAnti}\csentence{Robustness of network-based targeting. Shown is the final epidemic size (the percentage of individuals ever infected) for varying initial proportions of individuals holding pro-vaccine $\upsilon^{+}$ and anti-vaccine $\upsilon^{-}$, with the targeting method ${\cal O}_s$ set to {\it high-degree} in the top panel and {\it random} in the bottom. The insets show the case of $\upsilon^{+} = \upsilon^{-}$. Degree-based targeting provides more effective interventions than random placement. Each boxplot is based on 100 independent simulations.}}
\includegraphics[width=.9\columnwidth]{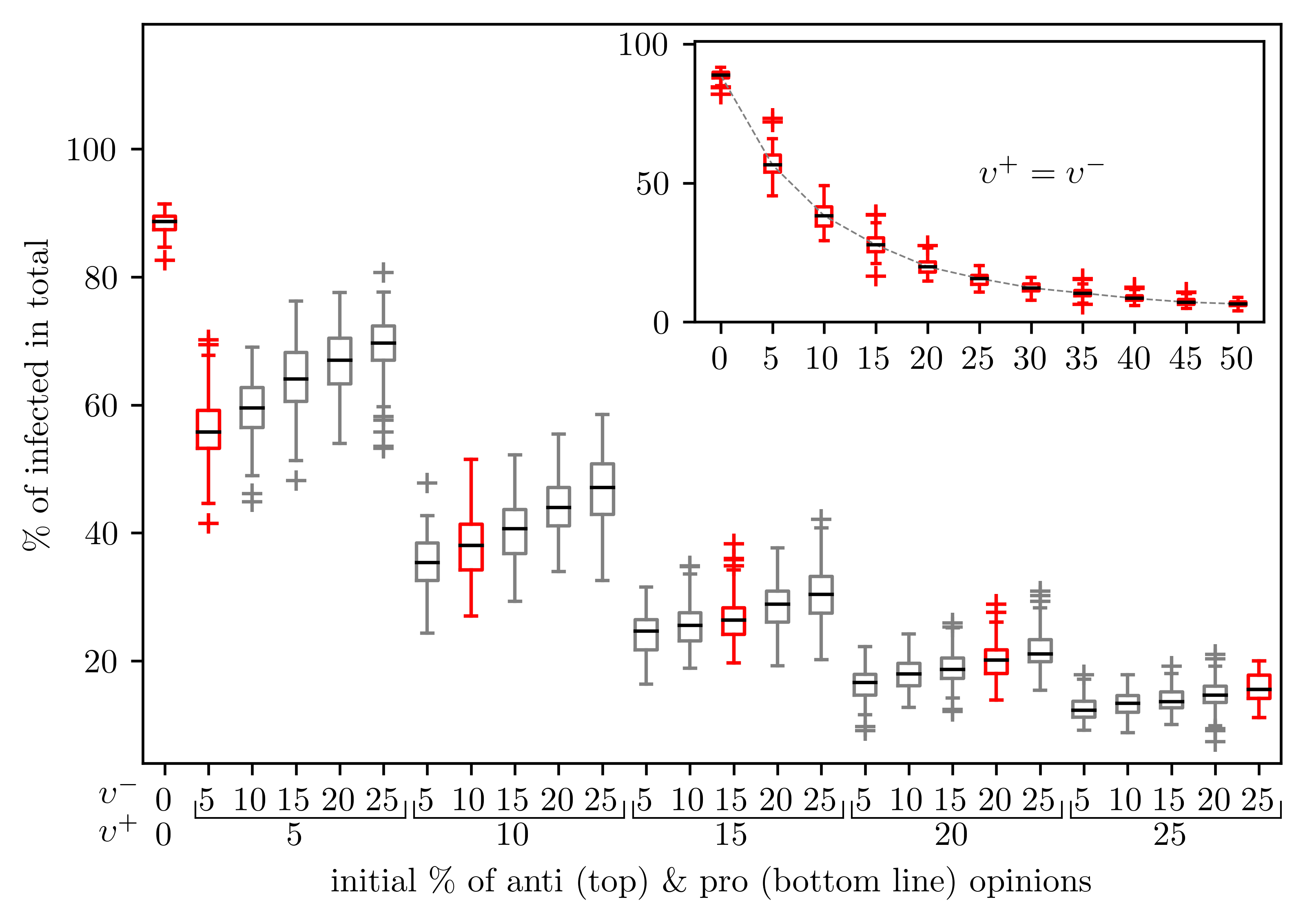}
\includegraphics[width=.9\columnwidth]{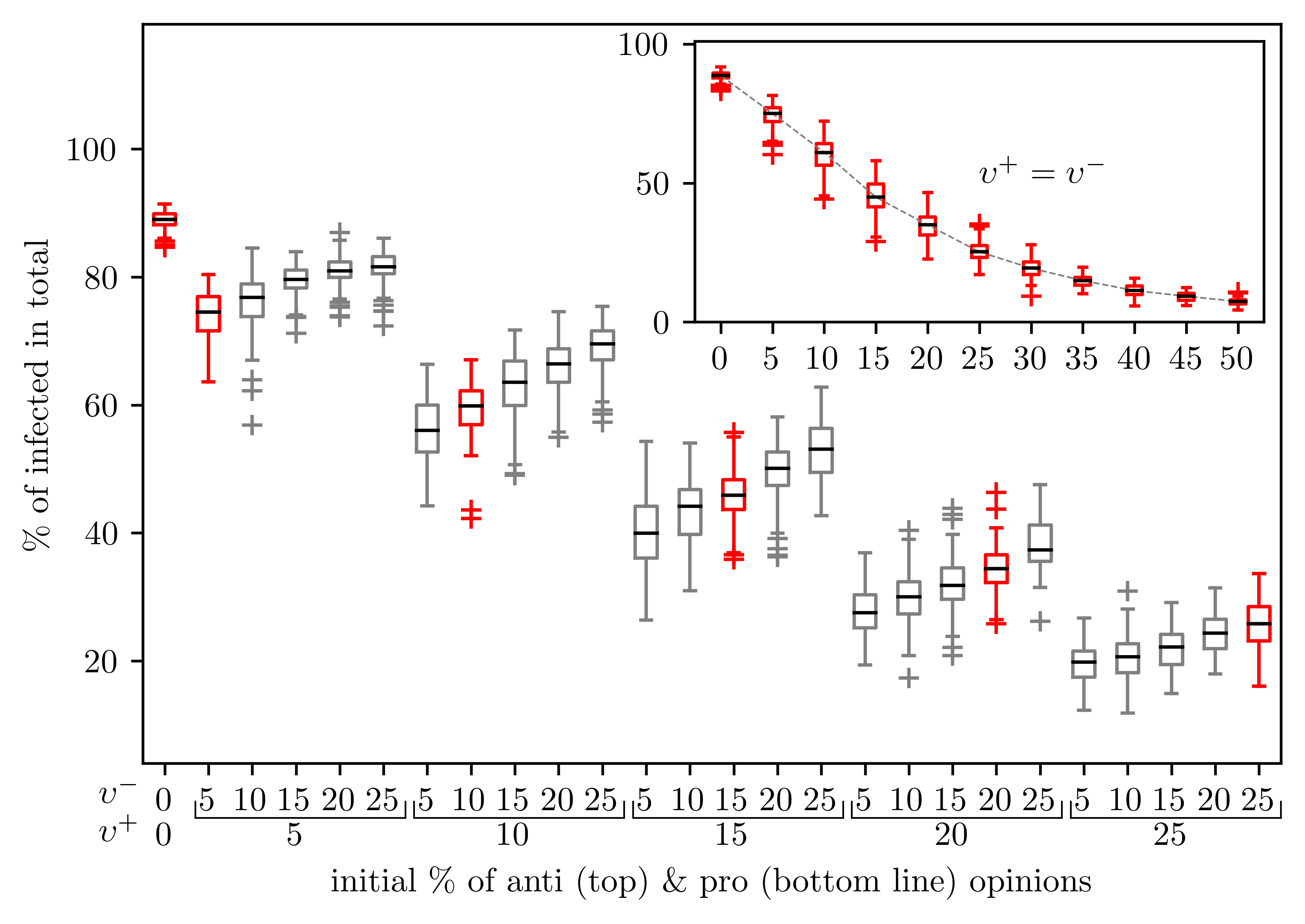}
\end{figure}

The outcome of an epidemic is shaped by the combination of the initial number of individuals holding pro-vaccine and anti-vaccine opinions at the outbreak. Specifically, we quantify the effect of these initial conditions on the infection curve by altering the parameters $\upsilon^{+}$ and $\upsilon^{-}$, which denote the percentage of individuals holding pro- and anti-vaccine opinions, respectively, at the beginning of each simulation. The top and bottom panels of Figure~\ref{fig:proAnti} depict the results of simulations with different targeting methods for the pro-vaccine opinion messengers, with ${\cal O}_s$ set to {\it high-degree} in the top panel and {\it random} in the bottom panel. All parameters other than $\upsilon^{+}$ and $\upsilon^{-}$ are fixed to their default values as specified in Table~\ref{tab:parameter}, and each simulation run starts with the generation of a fresh pair of networks for both layers.

In Figure~\ref{fig:proAnti}, each boxplot corresponds to a combination of initial conditions $\upsilon^{+}$ and $\upsilon^{-}$ with values taking from the set $\{5, 20, 15, 20, 25\}\%$, as indicated on the $x$-axis. On the $y$-axis, we have the total percentage of infected individuals (final epidemic size). As a comparison, we also show the base case where every individual has neutral attitude towards vaccine, that is, $\upsilon^{+}$ = $\upsilon^{-} = 0$. The insets of Figure~\ref{fig:proAnti} show how the final epidemic size decreases with an increasing equal presence of pro- and anti-vaccine individuals, including values of $\upsilon^{+}=\upsilon^{-}$ beyond $25\%$.

As expected, for a fixed proportion, $\upsilon^{+}$, of individuals seeded as pro-vaccine supporters, increases in anti-vaccine opinion individuals $\upsilon^{-}$ can cause greater final epidemic sizes, suggesting an approximately linear relationship with positive slopes.  However, a larger initial presence of pro-vaccine individuals $\upsilon^{+}$ leads to a greater extent of disease mitigation that strongly counteracts the negative impact due to the increase in  $\upsilon^{-}$. For example, for $\upsilon^{+} = 20\%$, the increase in total infections for bigger $\upsilon^{-}$ is limited, exhibiting a much smaller slope compared to the scenario $\upsilon^{+} = 5\%$. Regarding the symmetric initial conditions $\upsilon^{+} = \upsilon^{-}$, the final epidemic size decreases with increasing $\upsilon^{+} = \upsilon^{-}$; in particular, as $\upsilon^{+} = \upsilon^{-}$ increases beyond $30\%$, the spread of disease will be significantly suppressed. At $\upsilon^{+} = \upsilon^{-} = 50\%$ we see that essentially little or no disease outbreaks can unfold. These results hold for both targeting methods: {\it high-degree}-based and {\it random}. The former is much more effective than the latter, which is inline with Figure~\ref{fig:iniAdopt}. Overall, we confirm similar impact of varying initial conditions for other targeting methods ${\cal O}_s$.

\subsection*{Impact of network size}

\begin{figure}[h!]
\caption{\label{fig:gridsize}\csentence{Quantifying network size effect. Shown are the epidemic size (the percentage of individuals ever infected) (left panel) and the running times of the simulations (right panel) as a function of the total number of nodes in the network layers. Each boxplot is based on 100 independent simulations.}}
\includegraphics[width=.9\columnwidth]{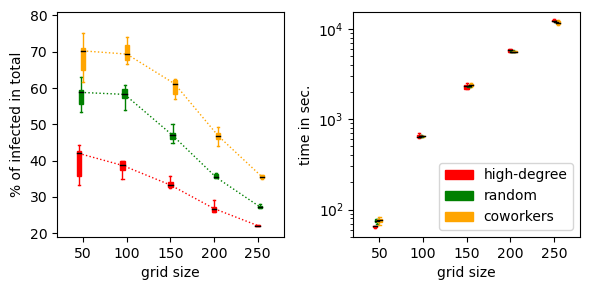}
\end{figure}

The previous simulations have been conducted with a fixed network size of $2500$ individuals. Figure~\ref{fig:gridsize} provides insight into the sensitivity of our simulation to larger network sizes. Beginning with the default $50\times50$ for the bottom layer of square lattice and top layer of Barab\`{a}si-Albert network, the remaining ticks on the $x$-axis represent network sizes of $10000$, $22500$, $40000$, and $62500$ nodes, respectively. To compare the impact of initial supporter placement methods (denoted as ${\cal O}_s$) in these test scenarios, we focus on three targeting algorithms ${\cal O}_s$, each of which corresponds to {\it high-degree}, {\it random}, and {\it coworkers}, as indicated by the legend in Figure~\ref{fig:gridsize}. Each boxplot summarizes the results of $10$ independent simulations.

\begin{figure}[h!]
\caption{\label{fig:times}\csentence{Comparing the running times of agent-based simulations with respect to specific implementations ${\cal O}_a$ of different targeting methods ${\cal O}_s$. This plot supplements the final results shown in Figure~\ref{fig:iniAdopt}. Each boxplot is based on 100 independent simulations.}}
\includegraphics[width=.9\columnwidth]{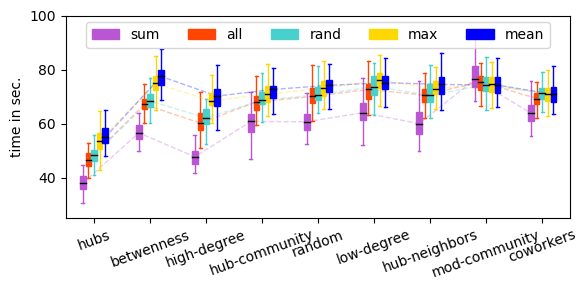}
\end{figure}

We observe an intriguing finite size effect: as the number of nodes in the layers increases, the final epidemic size decreases with smaller fluctuations, as shown in the left plot of Figure~\ref{fig:gridsize}. The relative position between the boxplots of different colors, representing different targeting methods, remains unchanged: {\it high-degree}-based targeting remains most effective in mitigating the disease impact, followed by {\it random} and then {\it coworkers}. This result further underscores the robustness of the simulation with regard to network size in comparing effectiveness of targeting methods. Overall, we confirm that the relative positions of the boxplots obtained for different targeting methods ${\cal O}_s$ as well as initial conditions of $(\upsilon^{-},\upsilon^{+})$-pairs remain qualitatively the same as in Figure~\ref{fig:iniAdopt} and~\ref{fig:proAnti}, respectively.

It is important to consider computing times when working with larger network sizes. The right plot in Figure~\ref{fig:gridsize} shows that the simulation times, which were conducted on a 3.2 GHz 16-Core Intel Xeon W with Turbo Boost up to 4.4 GHz and 768 GB RAM, increase steadily, approximately quadratically with the number of nodes. Starting with approximately one minute for a $50 \times 50$ square lattice/Barab\`{a}si-Albert network, the computation time can reach over three hours for a $250\times250$ network. We also find that the computation time is less sensitive to the specific targeting method ${\cal O}_s$.

Furthermore, we perform a comprehensive comparison of simulation times with respect to specific targeting methods ${\cal O}_s$ and opinion update methods ${\cal O}_a$ using the same default network size (Figure~\ref{fig:times}). Except for {\it betweenness}-based targeting, the computation times roughly track the trend of the final epidemic size (total infections) presented in Figure~\ref{fig:iniAdopt}. Generally and intuitively, the fewer infections there will be, the sooner the simulation will end.

\subsection*{Impact of average degree of the opinion layer}

\begin{figure}[h!]
\caption{\label{fig:ba}\csentence{Effect of varying average degree of the opinion network layer. Shown is the final epidemic size (percentage of individuals ever infected) for selected targeting methods (${\cal O}_s$ indicated on the $x$-axis) based on high degree nodes, random choosing, and community coworkers respectively with increasing network parameter, $m$. The average degree of the opinion layer is approximately $2m$, and thus larger $m$ values mean greater density of the network.}}
\includegraphics[width=.9\columnwidth]{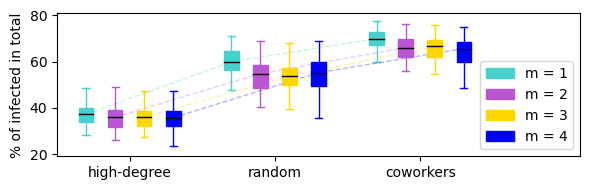}
\end{figure}

Lastly, we investigate the robustness of our simulation results with respect to varying the average degree of the opinion network layer. Keeping all other parameters fixed, we vary the number of edges added from a new node to existing ones (as indicated by the parameter $m$ in Table~\ref{tab:parameter}) from $m =1$ (average degree $\langle k \rangle = 2$) to $m = 4$ ($\langle k \rangle = 8$)  when generating the Barab\'{a}si-Albert network as the opinion formation layer in our model. As shown in Figure~\ref{fig:ba}, the effectiveness of disease mitigation by targeting high-degree, random, and coworkers nodes as initial pro-vaccine supporters remains largely unchanged. We also see that increasing the density of the opinion layer (via the parameter $m$) appears to yield noticeably yet insignificant lesser final epidemic size. Taken together, the density of the opinion layer has little, if any, impact on the social contagion process, whereas it is primarily the very structure of peer influence network and the network-based targeting methods that jointly play an important role in shaping the opinion propagation process.

\subsection*{Discussions and Conclusion}

In this work we investigate how various centrality-based targeting algorithms seeding the opinion influence network can impact the social contagion of pro-vaccine support as a means to control epidemic spreading on the other disease layer. While we consider a range of widely used opinion updating rules~\cite{sobkowicz2009}, it is helpful for future work to incorporate more realistic contagion models~\cite{dodds2004universal,campbell2013complex} and high-order interactions~\cite{iacopini2019simplicial,barrat2022social}, such as hypergraphs~\cite{barrat2022social}. The targeting algorithms also need to account for other factors, such as homophily (`birds of a feather to flock together')~\cite{centola2011experimental}, and the presence of central top-down campaign influence~\cite{wang2020public} apart from peer influence. By incorporating these extensions, we can enhance public confidence in vaccines at a time when coverage from child vaccinations to COVID-19 vaccinations may be declining or stagnant. 



The present work assumes a challenging scenario where the exchange and acceptance of views is much slower than disease spreading (using the time scale parameter $\omega=0.75$). It is worth noting that a prompt response from the population (smaller $\omega$ valuers) would create favorable conditions for mitigating the spread of disease~\cite{fu2017dueling}. Moreover, it is promising for future work to consider the impact of external shocks in the form of vaccine scares or skepticism, in order to overcome the hysteresis effect previously discovered in Ref.~\cite{chen2019imperfect}. The multiplex network targeting algorithms explored in this study can be further refined to identify individuals that are not only susceptible, but also responsive to, interventions. Keeping these in mind when optimizing network-based targeting methods, we will be able to harness the social contagion of vaccine knowledge and positive attitudes towards vaccination, with the goal of overcoming the hysteresis effect and increasing vaccination rates in areas of need. 

In conclusion, our results demonstrate that network-based targeting algorithms seeding the opinion layer with pro-vaccine supporters can greatly enhance attitude and behavior changes that are needed to control the spread of disease. Among those considered, targeting hubs -- individuals with the highest degrees in the influence network -- yields the most effective intervention that is not only able to flatten the curve with the smallest peak of infections but also able to curb the outbreak of the disease with the least number of total infections. The betweenness-based approach is similarly effective, but slightly lags behind the hub-based method. On the contrary, targeting groups of closely connected individuals either in the influence network ({\it mod-community}) or in the physical network ({\it co-workers}) leads to the least effective intervention due to the saturation effect, which is similarly observed in the field experiment~\cite{kim2015social}. Therefore, our study offers simulation-based insights to enhance the targeting effectiveness for future experiments.

\begin{backmatter}

\section*{Competing interests}
  The authors declare that they have no competing interests.


\section*{Acknowledgements}
F.F. gratefully acknowledges support from the Bill \& Melinda Gates Foundation (award no. OPP1217336), the NIH COBRE Program (grant no.1P20GM130454), and the Neukom CompX Faculty Grant. 

\bibliographystyle{bmc-mathphys} 
\bibliography{ans-references}      








\end{backmatter}
\end{document}